# Multiple-Fold Fermions and Topological Fermi Arcs Induced Catalytic Enhancement in Nanoporous Electride C12A7：4e⁻

Weizhen Meng, [1, 2] Xiaoming Zhang, *, [1, 2,] Ying Liu, [1, 2] Xuefang Dai [1, 2], Guodong Liu *, [1, 2], and Liangzhi Kou *, [3]

[1] State Key Laboratory of Reliability and Intelligence of Electrical Equipment, Hebei University of Technology, Tianjin 300130, China.
[2] School of Materials Science and Engineering, Hebei University of Technology, Tianjin 300130, China.
[3] School of Mechanical, Medical and Process Engineering, Queensland University of Technology, Garden Point Campus, QLD 4001, Brisbane, Australia
Email: zhangxiaoming87@hebut.edu.cn, gdliu1978@126.com, liangzhi.kou@qut.edu.au.

## Abstract

Topological materials are recently regarded as the idea catalysts due to the protected surface metallic states and high carrier mobility, however the fundamental mechanism and the underlying relationship between the catalytic performance and topological states are in debate. Here, by means of symmetry analysis and first-principles calculations, we discover that the electride material of C12A7:4e⁻ hosts the multiple-fold fermions due to the interstitial-electrons, with the sixfold- and fourfold- degenerate points locating at high symmetric points near the Fermi energy, which are identified as the underlying reason of the enhanced catalytic ability in C12A7:4e⁻-based catalysts. The multiple-fold fermions exhibit much longer Fermi arcs on the (001) surface than traditional Weyl/Dirac fermions, the surface is thus highly chemical active and possesses a low Gibbs free energy ($\Delta G_{H*}$ = 0.24 eV) for the hydrogen evolution reaction. The underlying relationship between catalytic performance and the topological surface state is explicitly verified by artificially hole doping, external strain and similar electride without the Fermi arcs, where the Gibbs free energies are significantly increased when the Fermi arcs is shifted to higher energy level. This work offers a guiding principle for understanding catalytic nature of electrides and the topological quantum catalysts.

**Keywords:** Electrides, Multiple-fold fermions, Long Fermi arcs, Topological quantum catalysts

# 1. Introduction

Topological materials [1-5], as novel types of quantum matter, have attracted considerable research interests currently. Early research on topological materials mainly focused on topological insulators with the insulating bulk state and conducting surface (or edge) states [1,2]. The research interests recently have been extended to topological semimetals (metals) [6] with carrying various topological fermions such as Weyl fermions [7-9], Dirac fermions [10-13], multiple-fold fermions [14-26], and nodal-line fermions [27-30], accompanied by novel physical characteristics. Especially, nontrivial surface states in topological materials have been hotly discussed to benefit the catalytic process due to their outstanding conductivity, high electron mobility and low Gibbs free energy [31-38]. Initially, such discussions were theoretically and experimentally propelled in topological insulators $Bi_2Se_3$ and $Bi_2Te_3$ [37,39,40], which observe the phenomenon of their catalytic enhancement in the fields of CO oxidation and hydrogen evolution. Soon after, Weyl semimetals in TaAs family were demonstrated to be effective in spurring the catalytic activity for hydrogen evolution reaction (HER) [31]. Quite recently, some other categories of topological semimetals were also proposed as potential catalysts, including high-topological-charge semimetals in PtAl, $PtSn_4$, $Nb_2S_2C$ and $V_{0.75}Ni_{0.25}Al_3$ and nodal-line semimetal in TiSi family [32-36]. These pioneering findings have strongly promised the feasibility of developing high-performance catalysts from topological semimetals. As an emerging field, the study on topological catalysts currently almost stays in superficial phenomenon discovery, however the fundamental mechanism and the underlying relationship between the catalytic performance and topological states are still unclear.

Electride is another special material category which has been evidenced as the platform of excellent catalysts for various heterogeneous catalysis reactions [41-44]. Electrides possess excess electrons, which are almost trapped in the interstitial positions of crystal lattice, serving as anions [45,46]. Due to the loosely bounded nature of the interstitial electrons, electrides usually exhibit high carrier mobility and low work function. These features arising from the presence of interstitial electrons have been long believed to be the nature for the high catalytic activity in electride-based catalysts. However, the mechanism is in debate from different experiments. A recent work [47] on electride $Ca_2N$ found the activity enhancement can be preserved even after the $Ca_2N$ is completely transformed into the $Ca_2NH$ hydride, where interstitial electrons of electride are annihilated. Another experimental work [48] on studying Ru/C12A7:4e⁻ (C12A7:4e⁻ represents $12CaO·7Al_2O_3$) catalyst, indicates that the surface-adsorbed hydrogen, rather than the hydride captured in the cages of electride $12CaO·7Al_2O_3$ (known as C12A7:4e⁻), is responsible for the high catalytic activity of reactive hydrogen in ammonia synthesis. Recently, some electrides are demonstrated to show intrinsic nontrivial band topology, termed as topological electrides [49-51]. To date, the band topology in topological electrides has been suggested to produce different topological phases such as topological insulators, Weyl/Dirac nodes, nodal lines and multiple-fold fermions [49-57]. Especially, some topological electrides were proposed to show floating surface bands [51], even different from traditional topological materials. These facts motivate us with that, topological electrides are naturally the ideal platform to clarify the underlying mechanism of catalytic enhancement in topological states and in electrides as well, because topological electrides simultaneously combines their both characteristics.

Motivated by above discussions, in present work we identified C12A7:4e⁻ as a new topological electride phase. We investigated the fundamental electronic structure of the material, and revealed the underlying relationship between the catalytic performance enhancement and topological surface states (extremely long Fermi arcs). It is found that this electride shows a nontrivial band topology, with coexisting of a sixfold- and a fourfold- degenerate points near the Fermi level. These multiple-fold fermions are mostly contributed by interstitial electrons confined inside the hollow cages. These symmetry-protected fermions characterize Fermi arc surface states, which almost traverse the entire surface Brillouin zone and are significantly longer than traditional Weyl/Dirac counterparts. Consequently, C12A7:4e⁻ harbors a highly active (001) surface. From the simulation of the inhibiting hydrogen poisoning (IHP) on the surface,

C12A7:4e⁻ exhibits the Gibbs free energy (ΔG_{H*}) as low as 0.24 eV, which is favorable for high-activity catalyst. The Gibbs free energy will be significantly increased when the Fermi arc surface states are shifted to high energy area or even removed by the hole doping, explicitly indicating the direct relationship between catalytic enhancement and topological surface state. Our work suggests that: during catalytic process in C12A7:4e⁻, although bulk cages themselves are not so active, their encaged interstitial electrons would contribute multiple-fold fermions, which activate the surface with high catalytic performance by producing giant surface Fermi arcs.

## 2. Calculation Methods

The first-principles calculations were performed in the framework of density functional theory (DFT) as implemented in the Vienna *ab*-initio simulation package (VASP). [58, 59] For electride C12A7:4e⁻, we used the generalized gradient approximation (GGA) of Perdew-Burke-Ernzerhof (PBE) method. [60] For electride C12A7:4e⁻, the cutoff value of plane wave kinetic energy was adopted as 400 eV, the Brillouin zone (BZ) was sampled with Γ-centered *k*-point mesh of 11×11×11 for both structural optimization and self-consistent calculations. The energy convergence criteria were chosen as $10^{-5}$ eV. For electride C12A7:4e⁻, the surface states were calculated by using Wannier functions and by using the iterative Green's function method as implemented in the WannierTools package.[61]

To obtain the Gibbs free energy ($\Delta G_{H*}$) of IHR on the (001) surface of C12A7:4e⁻, we adopt the method proposed by Norskov et al. [62, 63], the process can be written as:

$$H^+ + e^- \rightarrow H^{\#}. \quad (1)$$

Here, $H^{\#}$ is intermediate products. Then,

$$H^{\#} \rightarrow \frac{1}{2}H_2. \quad (2)$$

To get the $\Delta G_{H*}$ of the IHP on the [001] surface of C12A7:4e⁻, we first calculate the energy difference between the three terms as follows:

$$\Delta G_{H*} = \Delta E_H + \Delta E_{ZPE} - T\Delta S_H . \quad (3)$$

Here, $\Delta E_H$ is the adsorption energy, $\Delta E_{ZPE}$ is the change of the zero-point energy,

and $\Delta S_H$ is the change if the entropy of the adsorbed hydrogen on the [001] surface. First, $\Delta E_H$ can be written as:

$$\Delta E_H = [E_{([C12A7]+nH)} - E_{([C12A7])} + \frac{n}{2}E_{(H_2)}]. \quad (4)$$

Among, $E_{([C12A7]+nH)}$, $E_{([C12A7])}$, $E_{(H_2)}$ are the energies C12A7:4e⁻ with hydrogen adsorbed on (001) surface, the electride C12A7:4e⁻, and the $H_2$, respectively.

Second. $\Delta E_{ZPE}$ are 0.022 eV, which agrees well with previous calculated results. [62]

Finally, $\Delta S_H$ can be written as

$$\Delta S_H \approx -\frac{1}{2}S_H^0. \quad (5)$$

Here, $S_H^0$ is the entropy of the adsorption of hydrogen gas at the standard condition, [62] which is about 130.68 J/(K mol). [63]

Therefore, for electride C12A7:4e⁻, the $\Delta G_{H^*}$ of the IHP on the (001) surface can be written as:

$$\Delta G_{H^*} = \Delta E_H + 0.22 eV. \quad (6)$$

## 3. Results and discussion

### 3.1 Electride and topological electronic structure of C12A7:4e⁻

We start by charactering the crystal structure and the electride nature of C12A7:4e⁻. Figure 1(a) shows the unit cell of C12A7:4e⁻. It has a cubic structure, and the space group of $I\bar{4}3d$ (No. 220). C12A7:4e⁻ is a nanoporous material, the unite cell is constructed by 12 cages with large lattice space (~0.4 nm in diameter for per cage). As depicted in the lower corner of Fig. 1(a), the cage is composed with 6 Ca atoms, 7 Al atoms, and 16 O atoms. From the optimized crystal structure, the lattice constant is a= b= c= 10.48 Å. The Ca atoms occupy the 24*d* Wyckoff sites, Al atoms occupy two Wyckoff sites including 12*a* and 16*c*, and O atoms occupy Wyckoff sites 16*c* and 48*e*, respectively. Specific Wyckoff coordinates are listed in Table SI in section 1 the Supplementary Information. The optimized parameters well agree with the experiment values [64]. To show the electride nature of C12A7:4e⁻, we plot the electron localization function (ELF), a large region of electron is confined in the center of cages (the interstitial 12*b* Wyckoff site), as shown by the circled region of Fig. 1(b). The result confirms C12A7:4e¯ is indeed a typical 0D electride. In C12A7:4e¯, the Ca, Al, and O atoms show +2, +3, and -2 valence states respectively. As a result, the material totally contains 4 excess electrons per unit cell, and each cage shows the concentration fraction of 1/3 anionic electron. This scenario has been well identified by previous experiments [64].

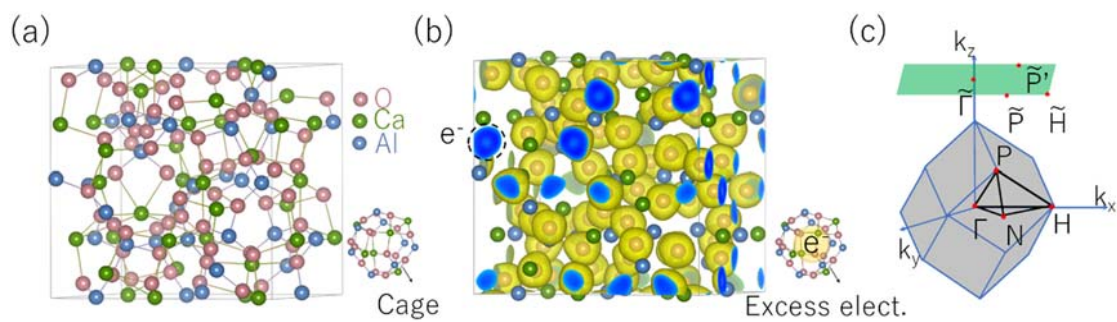

Fig. 1 (a) Crystal structure of electride C12A7:4e¯. The cage in C12A7:4e¯ lattice is shown in the lower corner of the figure. (b) Electron localization function (ELF) of electride C12A7:4e¯ with the isosurface values set as 0.65. The lower corner of (b) shows the confined electrons inside the cage. (c) Bulk and the (001) surface Brillouin zone of electride C12A7:4e¯ with high symmetry points indicated.

We then investigate the electronic structure of C12A7:4e¯ based on density functional theory (DFT) [65]. Since C12A7:4e¯ only contains light elements, the spin orbital coupling (SOC) is not included in the calculations since it has the trivial impact on the band structure, see section 2 of Supplementary Information. Figure 2(a) shows the band structure and partial density of states (PDOSs), C12A7:4e¯ is metallic. The band crossing points are located at the high symmetry points H and P near the Fermi level, as denoted by $P_1$ and $P_2$, respectively. From the PDOSs, we find the states near the Fermi level are mainly contributed by the interstitial electrons. This is also well consistent with the orbital component analysis for the bands near $P_1$ and $P_2$ [see the left panel of Fig. 2(b)]. In the right panel of Fig. 2(b), we show the partial electron density (PED) in the energy region of ±0.02 eV around $P_1$ and $P_2$. In these energy regions, large amounts of excess electrons are observed and locating inside the nanoporous cages.

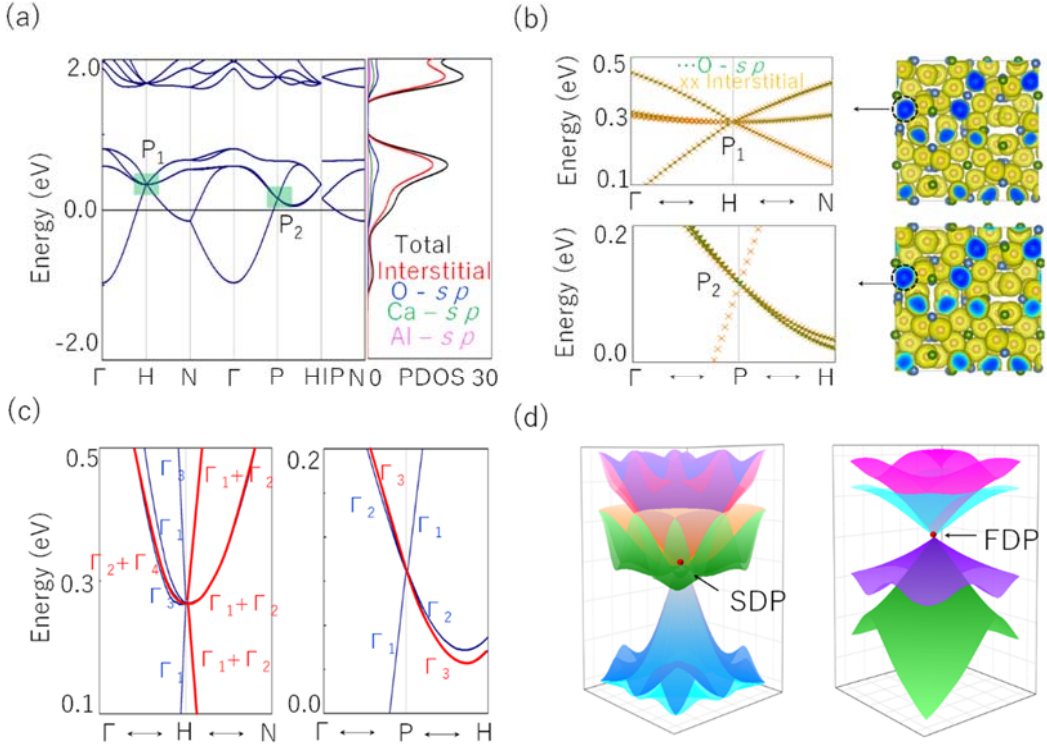

Fig. 2 (a) Band structure and partial density of states (PDOS) of electride C12A7:4e$^-$. The Fermi level is set as zero energy. The band crossings at the H and P points are denoted as $P_1$ and $P_2$. (b) The orbital-projected band structures near P1 and P2 are shown in the left panel. The right panel of (b) shows the partial electron density (PED) in the energy region of ±0.02 eV around $P_1$ and $P_2$. The isosurface value is chosen as 0.005 Bohr$^{-3}$. (c) The enlarged band structures along with the irreducible representations near $P_1$ and $P_2$. (d) The 3D plotting of band dispersions for the SDP and FDP.

Recently, an advanced work [66] has demonstrated the relations between the topological properties and electrides based on topological quantum chemistry theory [67]: the elementary band representations (BRs) can not only probe the nontrivial topology of electronic structures, but also identify the centers of electron density for specific bands (even for the floating bands from excess electrons). Based on this method, we perform the BRs analysis on the band crossings $P_1$ and $P_2$. The results are summarized in Table SI of section 1 of the Supplementary Information. Bands at $P_1$ and $P_2$ are both from the floating bands with BRs of A@12$b$, and the excess electrons inside the nanoporous cages (at the institute 12*b* Wyckoff sites). The findings fully agree with the results from DFT calculations in Fig. 2(b). The enlarged band structures near P$_1$ and P$_2$ along with the irreducible representations are displayed in Fig. 2(c). Detailed irreducible representations for other *k*-paths can be found in Table SII of section 3 of the Supplementary Information. Based on the symmetry analysis, we find P$_1$ at the H point is in fact a sixfold degenerate point (SDP), where the fatter bands are doubly degenerate while the thinner ones are singlets [see Fig. 2(c)]. The band crossing P$_2$ at the P point is a fourfold degenerate point (FDP). To be noted, the FDP in C12A7:4e$^-$ is fundamentally different from traditional Dirac points, which will be discussed later. The 3D plotting of band dispersions for P$_1$ and P$_2$ is shown Fig. 2(d).

We build effective model to further understand the nature of the SDP and FDP in C12A7:4e$^-$. From the symmetry analysis, the SDP at the H point, along the Γ-H path the bands split into four singlets with the irreducible representations of $\Gamma_3$, $\Gamma_1$, $\Gamma_3$, $\Gamma_1$ and one doubly degenerate band of $\Gamma_2 + \Gamma_4$ representation of the $C_{2v}$ symmetry [see Fig. 2(c)]. Along *k*-path H-N, the SDP splits into three doubly degenerate bands (nodal lines) with the $\Gamma_1 + \Gamma_2$ representation of the $C_s$ symmetry. These nodal lines are protected by the glide mirror symmetry $\widetilde{M}_{110}$ and the time-reversal symmetry (*T*). The little group at the H point belongs to $T_d$, which can be generated from $\{\widetilde{M}_{110} | \frac{1}{2}00\}$, $\{S_{4x}^{-1} | \frac{1}{2}00\}$, and $\{C_{3,1\bar{1}\bar{1}}^{-1} | 1\frac{1}{2}\frac{1}{2}\}$. By choosing proper basis, the effective Hamiltonian of the SDP taking the following form:

$$H_{SDP}(\boldsymbol{k}) = \begin{pmatrix} h_{11}(\boldsymbol{k}) & h_{12}(\boldsymbol{k}) \\ h_{21}(\boldsymbol{k}) & h_{22}(\boldsymbol{k}) \end{pmatrix}. \quad (7)$$

Here, $h_{ij}(\boldsymbol{k})$ is a $3 \times 3$ matrix. Specifically,

$$h_{11}(\boldsymbol{k}) = v\boldsymbol{k} \cdot \boldsymbol{S}, h_{12}(\boldsymbol{k}) = -v\boldsymbol{k} \cdot \boldsymbol{S}, \quad (8)$$

$$h_{12}(\boldsymbol{k}) = h_{21}^*(\boldsymbol{k}) = \begin{pmatrix} 0 & -\gamma' k_x & \gamma'' k_y \\ -\gamma'' k_x & 0 & \gamma' k_z \\ \gamma' k_y & \gamma'' k_z & 0 \end{pmatrix} \quad (9)$$

Here, $v$ is a real parameter. If we consider a limitation that $|v| \gg |\gamma'^{(\prime\prime)}|$, such an SDP carries a zero Chern number which is a composition of two spin-1 fermions. Here, we want to propose that, the SDP can be an excellent platform to realize other fermions such as eightfold-, fourfold-, and threefold fermion under SOC or lattice strain (see section 2 and 4 of Supplementary Information).

For the FDP at the P point [see Fig. 2(c)], the FDP along the Γ-P-H path splits into two singlets and one doubly degenerate band, which belong to the irreducible representations of $\Gamma_1$, $\Gamma_2$, and $\Gamma_3$ of the $C_{3v}$ symmetry, respectively. The FDP at the P point is protected by the $C_{3v}$ symmetry, which can be treated with four independent generators including $\{S_{4x}^+ | \frac{1}{2}00\}$, $\{C_{3,\bar{1}11}^+ | 0\frac{1}{2}\frac{1}{2}\}$, $\{C_{2y} | 0\frac{1}{2}\frac{1}{2}\}$, and $\{C_{2x} | \frac{3}{2}\frac{3}{2}0\}$. Based on them, the effective Hamiltonian up to the first order can be generally expressed as,

$$H_{FDP}(\boldsymbol{k}) = \begin{pmatrix} h_{11}(\boldsymbol{k}) & h_{12}(\boldsymbol{k}) \\ h_{21}(\boldsymbol{k}) & h_{22}(\boldsymbol{k}) \end{pmatrix}, \quad (10)$$

In details, each $h_{ij}(\boldsymbol{k})$ is a $2 \times 2$ matrix which can be given as,

$$h_{11}(\boldsymbol{k}) = v(k_x + ik_y)\sigma_+ + h.c., \quad h_{22}(\boldsymbol{k}) = -h_{11}(\boldsymbol{k}), \quad (11)$$

$$h_{12}(\boldsymbol{k}) = e^{i\theta_1} A k_z \sigma_z + [e^{i\theta_2} B(k_x + ik_y)\sigma_+ + h.c.], \quad h_{21}(\boldsymbol{k}) = h_{12}^*(\boldsymbol{k}).$$

Here, $\sigma$'s is the Pauli matrix, $\sigma_\pm = (\sigma_x \pm i\sigma_y)/2$. $v = 3\alpha + \beta$, $A = \sqrt{(\alpha+\beta)^2 + 4\alpha^2}$, $B = \sqrt{2}\sqrt{4(\alpha-\beta)^2 + v^2}/4$, the parameters $\alpha$ and $\beta$ are real values which can be derived from DFT calculations. $\theta_1$ and $\theta_2$ are also real values, $\theta_1 = \arg[(\alpha+\beta) - i2\alpha]$, $\theta_2 = \arg[-2(\alpha-\beta) + i(3\alpha+\beta)] + \pi/4$.

The models provided above fully clarify from the symmetry view the presences of SDP and FDP in C12A7:4e⁻, and more importantly, suggest that these multiple-fold fermions are robust against weak perturbations that preserve the symmetry. This fact enables us to removed unconcerned characteristics of C12A7:4e⁻ with the multiple-fold fermions preserved via applying weak perturbations, such as hole doping and hydrostatic lattice distortion. This is important to investigate the catalytic nature in C12A7:4e⁻, which will be fully discussed later.

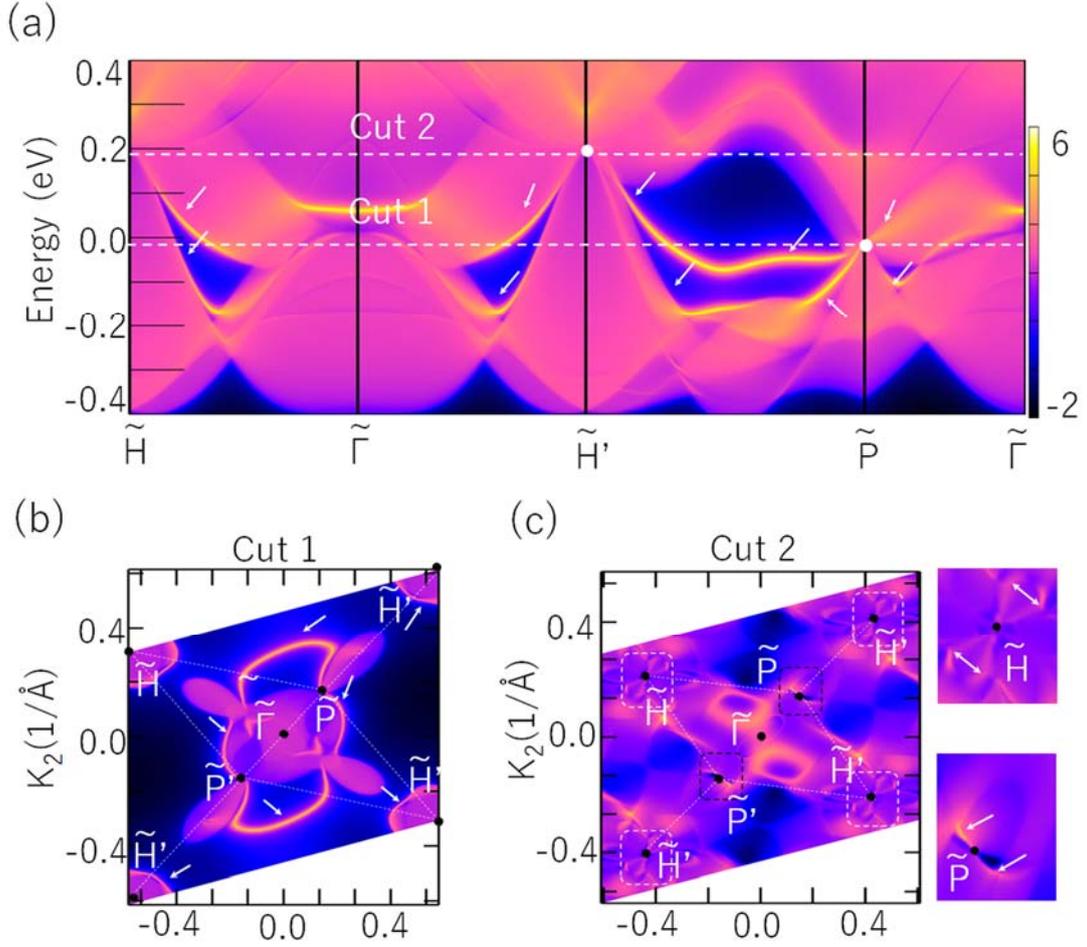

Fig. 3 (a) shows the (0 0 1) surface states of electride C12A7:4e⁻. (b) and (c) show the (001) surface slice under cut 1 (at energy of -0.01 eV) and cut 2 (0.19 eV), respectively. In (a), (b), (c), the Fermi arcs are pointed by the write arrows. In (c), the illustrations on the right panel are the partial enlarged view of the states in the white and black boxes.

**3.2 Surface properties and catalytic activity of C12A7:4e⁻**

From the above analysis, C12A7:4e⁻ is clearly identified as a topological electride with coexisting of SDP and FDP fermions. Here we investigate the potential surface states of these fermions. Figure 3(a) shows the (001) surface states along the surface paths shown in Fig. 1(c). Two pieces of Fermi arcs originating from the $\widetilde{H}$ ($\widetilde{H'}$) and $\widetilde{P}$ points are observed. From the surface slices at the SDP and FDP fermions, it is found that the arcs of SDP show the dumbbell-like shape, and cover most region of (001) surface [see Fig. 3(a) and (c)]. In contrast, two pieces of Fermi arcs of FDP are in the petal-like shape [see Fig. 3(a) and (b)], which are extremely long, nearly traversing the entire surface Brillouin zone.

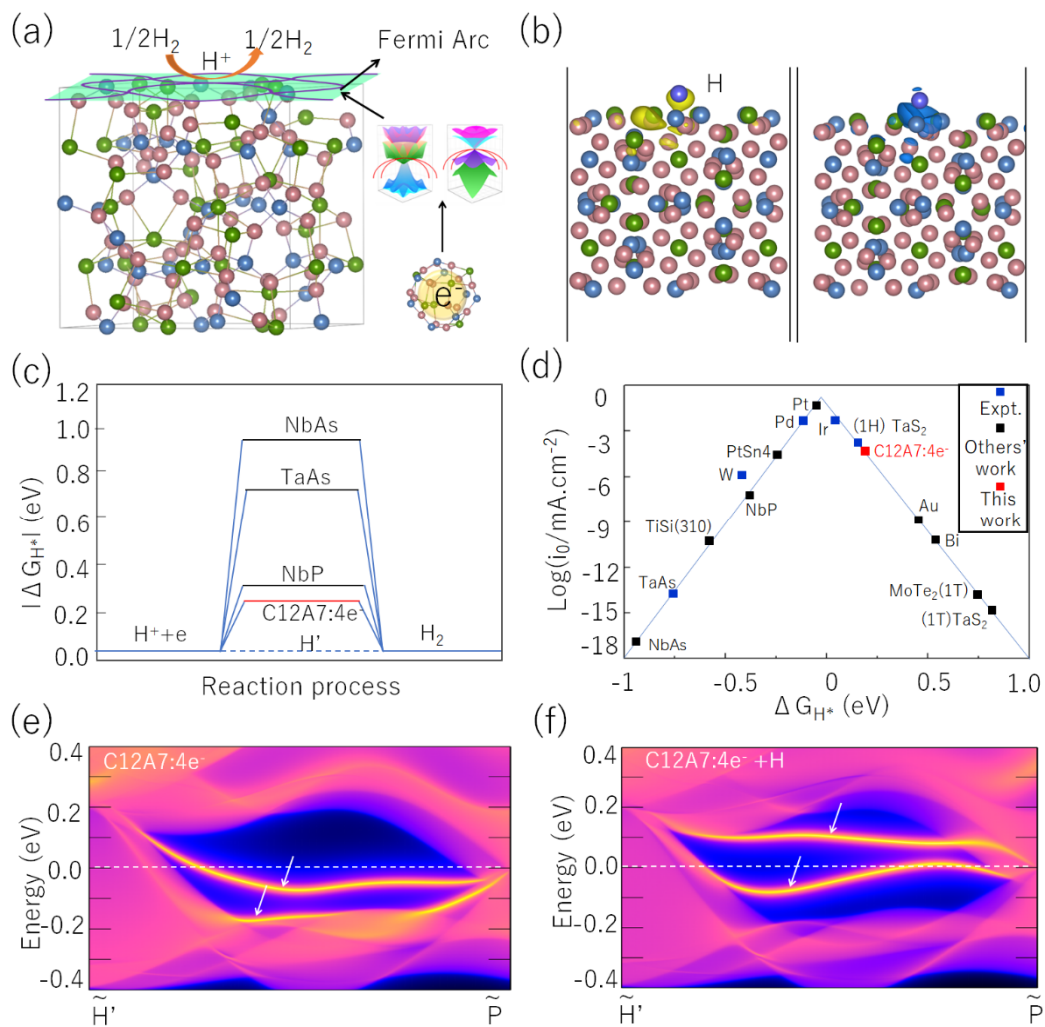

Fig. 4 (a) Schematic of the mechanism for IHP process on the (001) surface of C12A7:4e⁻: the long Fermi arcs induced by the SDP and FDP excitations activate the surface for IHP reaction. (b) The electron depletion (the left pannel) and accumulation (the right panel) during H adsorption on C12A7:4e⁻ surface. The isosurface values are set to 0.0025 e Å⁻³. (c) The free energy diagram for hydrogen evolution at a potential $U = 0$ relative to the standard hydrogen electrode at pH = 0. The free energy of $H^+ + e^-$ is by definition the same as that of $1/2\ H_2$ at standard condition of equilibrium. The data for NbP, TaAs and NbAs are taken from ref. 31. (d) Volcano plot for IHP of C12A7:4e⁻ in comparison with typical catalysts. The data are taken from refs. 31, 38, 62, 69. (e) and (f) are the (001) surface states at specific path of C12A7:4e⁻ without and with the hydrogen adsorption, respectively.

Similar to traditional Weyl/Dirac topological quantum catalysts such as TaAs and PtAl [31,32], it is confirmed that C12A7:4e$^-$ naturally possesses topological bulk band structure and nontrivial surface states. However, comparing with these examples, C12A7:4e$^-$ is different in two aspects: (1) C12A7:4e$^-$ is an electride, and the fermions are contributed by the excess electrons in the interstitial cites which are more loosely bounded, (2) C12A7:4e$^-$ possesses multiple-fold fermions (SDP and FDP), which show significantly longer Fermi arcs than Weyl/Dirac counterparts. These features are expected to play a significant role in the enhancement of the chemical activities as the topological quantum catalysts.

In the following, we therefore study the catalytic performance of the material with and without the topological surface states, by estimating the adsorption of H and examining the chemical process of H$_2$ production on the (001) surface of C12A7:4e$^-$, as shown in Fig. 4(a). This corresponds to the IHP process for ammonia synthesis. Here, we want to point out that, the IHP process is actually similar to the hydrogen evolution reaction (HER) process discussed in typical topological quantum catalysts TaAs and PtAl [31, 32]. Here we use the $\Delta G_{H*}$ of hydrogen adsorption on the catalyst surface to judge the IHP activity. As expected, C12A7:4e$^-$ indeed shows a relatively low $\Delta G_{H*}$ of 0.24 eV. As shown in Fig. 4(c), the absolute value $|\Delta G_{H^*}|$ in C12A7:4e$^-$ is significantly lower than those in Weyl catalysts NbP (0.31 eV), TaAs (0.74 eV) and NbAs (0.96 eV) [31]. Moreover, we plot the volcanic curves and compare the IHP performance of electride C12A7:4e$^-$ with typical catalysts and topological quantum catalysts. As shown in Fig. 4(d), we find C12A7:4e$^-$ nearly sits at the top of the volcanic curves, indicating the highly active surface. In Fig. 4(e) and (f), we further compare the surface states of C12A7:4e$^-$ before and after hydrogen adsorption, and find the long Fermi arcs are robust during the IHP process, but shift upward due to the electron donation from the *s* orbital of the adsorbed hydrogen. This process can be further confirmed from the charge density difference (CDD) in Fig. 4(b), the charge depletion occurs on the surface of C12A7:4e$^-$, while charge accumulation happens around the adatom.

## 3.3 Evidence of topologically catalytic enhancement

To clarify the role of the topological fermions on the excellent catalytic performance, we calculated the Gibbs free energies by artificially removing or shifting the Fermi arcs. First of all, we estimate the catalytic activity in the bulk, especially in the bulk lattice cages. Our results show $\Delta G_{H*}$ inside the bulk cages is remarkably larger (absolute value: 0.51 eV) than that on the surface (0.24 eV). It is thus clear that the surface of C12A7:4e⁻ is much more active for catalysis than the nano-sized cages inside the bulk. These results can give a reasonable explanation to the recent experimental findings of the enhanced catalytic performance in Ru/C12A7:4e⁻ [48], which show the bulk cages of C12A7:4e⁻ themselves are not active during the catalytic process. From our results, although the bulk cages are not active, the encaged excess electrons can contribute multiple-fold fermions in the bulk electronic state. These fermions further produce extremely long Fermi arcs on the surface, making the surface states highly active which are beneficial for the chemical catalytic process. This mechanism has been conceptually shown in Fig. 4(a).

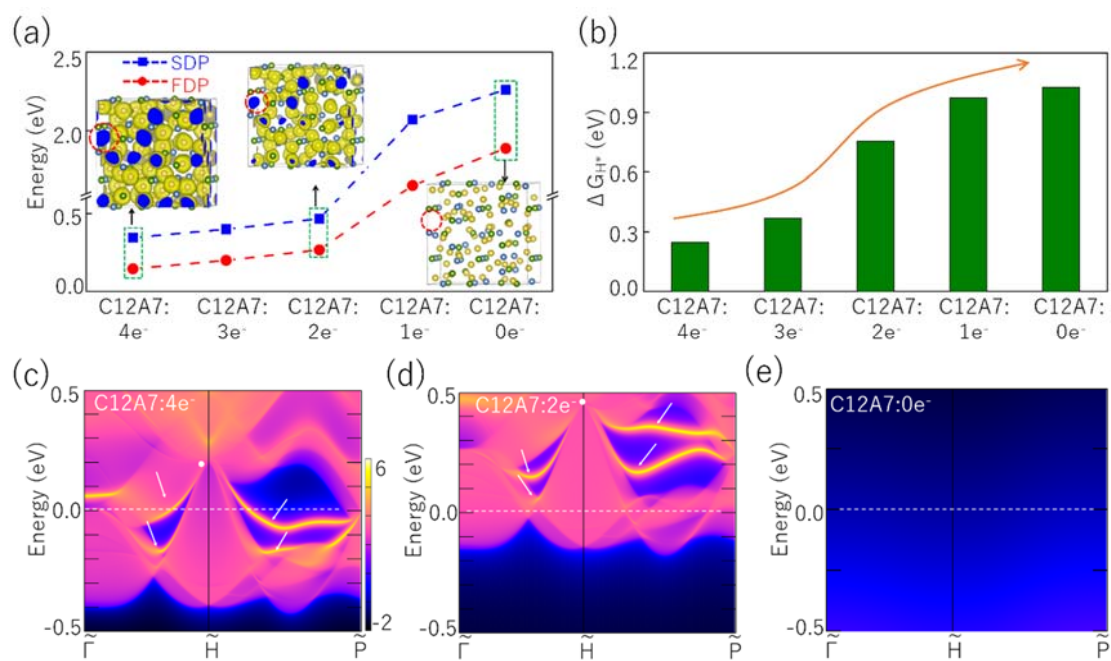

Fig. 5 (a) The positions of SDP and FDP in C12A7:4e⁻, C12A7:3e⁻, C12A7:2e⁻, C12A7:1e⁻, C12A7:0e⁻. The insets in (a) provide the ELF maps for C12A7:4e⁻, C12A7:2e⁻, and C12A7:0e⁻. (b)

The (001) surface Gibbs free energy in C12A7:4e⁻, C12A7:3e⁻, C12A7:2e⁻, C12A7:1e⁻, C12A7:0e⁻. (c), (d) and (e) are the (001) surface states at specific paths for C12A7:4e⁻, C12A7:2e⁻, C12A7:0e⁻, respectively. In (c) and (d), the Fermi arcs are pointed by the write arrows.

On the other hand, we investigated the catalytic activities on the (001) surface after removing the effects of the bulk multiple-fold fermions and surface Fermi arcs in C12A7:4e⁻. This scenario can be realized by artificially annihilating the excess electrons (via hole doping) in C12A7:4e⁻. As discussed above, one cell of C12A7:4e⁻ contains 4 excess electrons, which are almost localized in the interstitial space of the cages. By annihilating 1, 2, and 3 electrons in the system, we find the localized electrons in the cages are reduced. After 4 excess electrons being fully annihilated, the system is no longer an electride phase without containing any caged electrons. This evolution process can be clearly observed in the ELF maps, as shown by the insets of Fig. 5(a). To be noted, as indicated by the effective models displayed earlier, the SDP and FDP will not annihilate during the period because the system symmetry is unchanged. However, associated with the reduced excess electrons, the SDP and FDP will move away from the Fermi level towards higher energy levels, as shown in Fig. 5(a). The specific band structures for all the cases are provided in the section 5 of Supplementary Information. Correspondingly, as shown in Fig. 5(b), the $\Delta G_{H^*}$ value of the system increases with the SDP and FDP moving away from the Fermi level, suggesting the high dependence of catalytic activity on the surface Fermi arcs during the period. Such reduction of the catalytic performance can be contributed to the shift of topological surface states in the system. In Fig. 5(c)-(e), we compare the surface states within $|E-E_F| = 0.5$ eV for native C12A7:4e⁻, and the cases with annihilating 2 (C12A7:2e⁻) and 4 (C12A7:0e⁻) excess electrons. One can clearly see sufficient surface Fermi arcs at the Fermi level for native C12A7:4e⁻. In C12A7:2e⁻, although surface Fermi arcs exist, all the regions of fermi arcs unfortunately locate above the Fermi level. For C12A7:0e⁻, no conductive surface states can be found in the energy region of ±0.5 eV. The surface activity in these systems follows the order of: C12A7:4e⁻ > C12A7:2e⁻ > C12A7:0e⁻.

This well agrees with the obtained ΔG$_{H*}$ in Fig. 5(b), which follows: C12A7:4e⁻ < C12A7:2e⁻ < C12A7:0e⁻. The results have explicitly demonstrated that the excellent catalytic performance is originated from the multiple-fold fermions and corresponding long surface Fermi arcs near the Fermi level in electride C12A7:4e⁻.

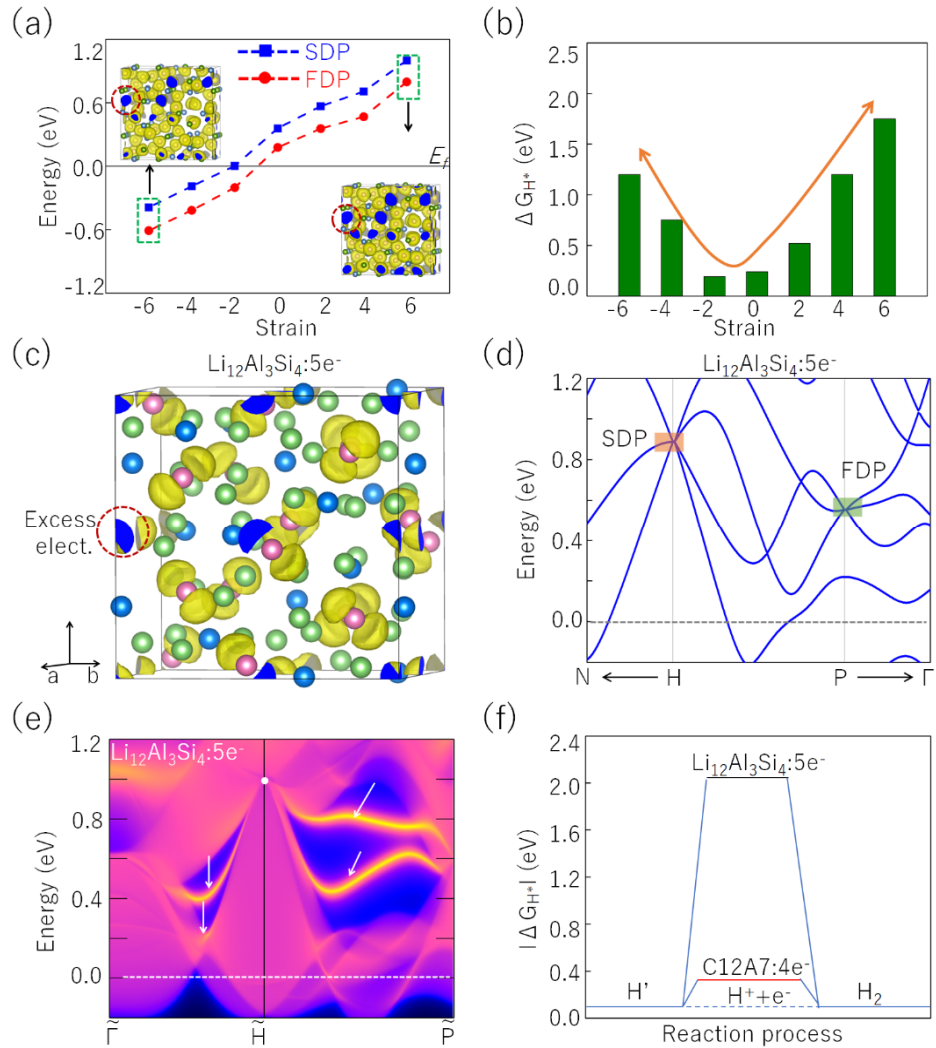

Fig. 6 (a) The energy position of SDP and HDP fermions corresponding to the Fermi level ($E_f$) under hydrostatic distortions from -6% to +6%, where negative and positive values denote lattice compression and lattice expansion, respectively. The insets in (a) provide the ELF maps of electride C12A7:4e⁻ under -6% and +6% lattice distortions. (b) The change of ΔG$_{H*}$ on the (001) surface of C12A7:4e⁻ under different lattice distortions. (c) The ELF map of electride Li$_{12}$Al$_3$Si$_4$:5e⁻ with the isosurface values set as 0.8. (d) The electronic band structure for electride Li$_{12}$Al$_3$Si$_4$:5e⁻. (c) Volcano plot for IHP of electride Li$_{12}$Al$_3$Si$_4$:5e$^-$ in comparison with electride C12A7:4e$^-$. (f) The (001) surface states at specific paths for Li$_{12}$Al$_3$Si$_4$:5e$^-$. The Fermi arcs are pointed by the write arrows.

Furthermore, to verify the catalytic enhancement of electride C12A7:4e$^-$ arises from surface Fermi arcs rather than from the excess electrons themselves, we make further discussion on two aspects. First, we perform hydrostatic distortions to the C12A7:4e$^-$ lattice. The distortion is arranged from -6% to +6%, where negative and positive values denote lattice compression and lattice expansion, respectively. As shown the ELF maps in Fig. 6(a) (also in Fig. S6 of Supplementary Information), we find the electride feature in C12A7:4e$^-$ is nearly unchanged during the lattice distortion. However, we find the positions of the multiple-fold fermions are extremely sensitive to the lattice distortion. That is, the SDP and FDP will move away from the Fermi level regardless of lattice compression and lattice expansion, as shown in Fig. 6(a). The specific band structures for all the cases have been provided in Fig. S5 of Supplementary Information. For these cases, we calculate the $\Delta G_{H*}$ for the (001) surface under different lattice distortions, as shown in Fig. 6(b). Interestingly, we find the $\Delta G_{H*}$ are relative to the distance of SDP/HDP fermions corresponding to the Fermi level. That is, the $\Delta G_{H*}$ tends to the zero energy if the Fermions are close to the Fermi level. These results fully indicate that the catalytic enhancement for electride C12A7:4e$^-$ originates from the multiple-fold-fermions-induced Fermi arc on the (001) surface.

For the other aspect, we have identified another electride phase namely Li$_{12}$Al$_3$Si$_4$:5e$^-$, which follows the same crystal space group with C12A7:4e$^-$. The specific crystal structures, valence state analysis and atomic positions have been provided in the section 7 of Supplementary Information. In Fig 6(c), we display the ELF map of Li$_{12}$Al$_3$Si$_4$, where the excess electrons are trapped in interstitial sites (see the red circle). The electronic band structure of electride Li$_{12}$Al$_3$Si$_4$:5e$^-$ is shown in Fig. 6(d). Since the symmetry operations are the same with C12A7:4e$^-$, Li$_{12}$Al$_3$Si$_4$:5e$^-$ also show the SDP and FDP on the high symmetry points H and P, respectively. However, unlike electride C12A7:4e¯, the FDP and FDP in electride Li$_{12}$Al$_3$Si$_4$:5e$^-$ are far away from the Femi level (above 0.6 eV). These fermions and their surface states are unlikely contribute to the conducting active in the system, as shown in Fig. 6(e). Barely with the electride contribution, Li$_{12}$Al$_3$Si$_4$:5e$^-$ shows a much higher $\Delta G_{H*}$ on the (001) surface than C12A7:4e¯ (1.90 eV *versus* 0.24 eV), as shown in Fig. 6(f).

Our research work for the first time explicitly revealed the topological surface Fermi arcs and its important role in catalytic chemical reactions, which provide deep understanding for topological catalysis and guidance for experimental debates.

### 4. Summary

In summary, from first principle simulations we revealed the presences of nontrivial band topology and multiple-fold fermions (including a SDP and a FDP) in electride C12A7:4e¯. The multiple-fold fermions are formed by interstitial-electrons in the electride, which are distinct from the previous topological materials. The fermions are protected by symmetry, and show novel surface states with long Fermi arcs nearly traversing the entire (001) surface Brillouin zone. The long Fermi arcs originating from the multiple-fold fermions make the C12A7:4e¯ (001) surface highly active, and lead to a relatively low $\Delta G_{H*}$ (0.24 eV) during the IHP catalytic process. Notably, this $\Delta G_{H*}$ is lower than most topological quantum catalysts proposed previously. Our work provides new understanding of the catalytic nature in C12A7:4e¯-based catalysts by combing the topological quantum states.

**Conflicts of interest**

There are no conflicts of interest to declare.

**Acknowledgements**

This work is supported by National Natural Science Foundation of China (Grants No. 11904074). The work is funded by Science and Technology Project of Hebei Education

Department, the Nature Science Foundation of Hebei Province (Nos. A2019202222 and A2019202107), the Overseas Scientists Sponsorship Program by Hebei Province (C20200319). The work is also supported by the State Key Laboratory of Reliability and Intelligence of Electrical Equipment (No. EERI_OY2020001), Hebei University of Technology. One of the authors (X.M. Zhang) acknowledges the financial support from Young Elite Scientists Sponsorship Program by Tianjin.